\documentclass[12pt,preprint]{aastex}

\shorttitle{}
\usepackage{graphicx}% Include figure files
\usepackage{bm}% bold math
\usepackage{amssymb}

\begin{document}


\title{Effect of a non null pressure on the evolution of perturbations in the matter dominated epoch}


\author{A. Herrero}
\affil{Departament de Matem\`atica Aplicada, Universitat Polit\`ecnica de
       Val\`encia, Spain}
\email{aherrero@mat.upv.es}

\and 

\author{M. Portilla}
\affil{Departament d'Astronomia i Astrof\'{\i}sica, Universitat de
       Val\`encia, Spain}
\email{miguel.portilla@uv.es}


\begin{abstract}
We analyze the effect of pressure on the evolution of perturbations of an
Einstein-de Sitter Universe in the matter dominated epoch assuming an ideal
gas equation of state. For the sake of simplicity the temperature is
considered uniform. The goal of the paper is to examine the validity of the
linear approximation. With this purpose the evolution equations are developed
including quadratic terms in the derivatives of the metric perturbations and
using coordinate conditions that, in the linear case, reduce to the
longitudinal gauge. We obtain the general solution, in the coordinate space,
of the evolution equation for the scalar mode, and, in the case of spherical
symmetry, we express this solution in terms of unidimensional integrals of
the initial conditions: the initial values of the Newtonian potential and its
first time derivative. We find that the contribution of the initial first
time derivative, which has been systematically forgotten, allows to form
inhomogeneities similar to a cluster of galaxies starting with very small
density contrast. Finally, we obtain the first non linear correction to the
linearized solution due to the quadratic terms in the evolution equations.
Here we find that a non null pressure plays a crucial role in constraining
the non linear corrections. It is shown, by means of examples, that
reasonable thermal velocities at the present epoch (not bigger than $10^{-6}$)
make the ratio between the first non linear correction and the linear
solution of the order of $10^{-2}$ for a galaxy cluster inhomogeneity.

\end{abstract}


\keywords{cosmology: theory, galaxies: clusters: general}


\section{Introduction}

The relativistic theory of the evolution of perturbations was initiated in
1946  by Lifshitz using a special coordinate condition known as the
synchronous gauge. He linearized the Eintein's equations to obtain the
evolution of perturbations. In fact, he found plane wave solutions for the
radiation dominated epoch, assuming $p= (1/3) \rho$ as equation of state,
and for the matter dominated epoch, neglecting the effects of the pressure.
The theory, with subsequent improvements, is referenced in many books of
Cosmology \citep{PEEB,ZEL,LAN,WEI}.

However, the synchronous coordinate condition has two great drawbacks. The
first one is consequence of the fact that it does not completely fix the
coordinate system, allowing the existence of gauge modes. This problem can be
handled using a gauge invariant version of the theory, started by Bardeen
\citep{BAR} and collected by Mukhanov, Feldman \& Branderberger \citep{MUK}.
This last review also shows an easier way to obtain gauge invariant equations
using a coordinate condition that does not allow the existence of gauge modes.
These coordinate conditions define what is known as the longitudinal gauge.

The other inconvenience of the synchronous gauge is that the metric
perturbation and the density contrast both depend on the second space-like
derivatives of a potential. Then, great values for the density contrast imply
great values for the metric perturbation, and in consequence the linear
approximation in this gauge fails when the density contrast is bigger than
unity. On the contrary, in the longitudinal gauge the metric perturbation is
proportional to a potential while the density contrast is proportional to the
laplacian of the same potential. So, the metric perturbation can be a very
small quantity while the corresponding density contrast can achieve values
greater than unity. For example, galaxy clusters develop a potential of the
order of $\phi/c^2 \leq  10^{-5}$ varying at scales of $R\approx 1Mpc/6000 h$;
then, using the relation $\delta =(1/6) \Delta \phi$ for the density contrast,
in adimensional coordinates, we get $\delta \approx 1/(6 R^2) \approx  10^3$
for $h=0.5$. This makes, in principle, possible the validity of the linear
approximation in the longitudinal gauge to study the formation of
inhomogeneities similar to galaxy clusters.

Then, the question arises why the linear approximation (linear in the
metric perturbation) is always considered inaccurated to describe the
evolution when the density contrast is bigger than unity. In this paper we
are interested in analyzing when the linear approximation begins to fail in
describing the evolution of such objects. To do that, according to the
previous paragraph, it is necessary to take into account here on that the
spatial derivatives of the potential can be much bigger than the potential.
Let us to point, in advance, that the pressure plays a crucial role in this
issue.

So, in section \ref{EV} we write Einstein's equations in evolutive form,
keeping quadratic terms in the first derivatives of the metric perturbation
and neglecting quadratic terms in the potential. We use coordinates which
simplify the evolution of the tensor components of the metric perturbation
and which reduce, in the linear case, to the longitudinal gauge. Moreover, to
complete the evolution equations we need to give the stress tensor. In the
matter dominated epoch and after the decoupling with radiation, the
temperature of matter $T(t)$ decreases as $1/a^2(t)$, where $a(t)$ is the
expansion factor, and the pressure becomes so small that it is usually
neglected. But, as we shall see in this paper a non null pressure is necessary
to keep valid the linear approximation. Then, we will consider the simple case
of an ideal gas with an equation of state of the form $p=(a_o^2 T_o/ma^2(t))
\rho$. Notice that, although the temperature is becoming very small, the 
evolution will increase the energy density and this is the reason for keeping
the pressure. Under these conditions we write down, in section \ref{LEE}, the 
evolution equations of the gravitational potential in the linear approximation.

The evolution equations have two degrees of freedom: the potential and its
first time derivative. In section \ref{CPLE} we find the general solution
$\phi^{(0)}$ of the linear evolution equations for arbitrary initial
conditions. Next, we use this result in section \ref{EFNN} to obtain the
evolution of the density contrast and the macroscopic velocity starting from
appropriated initial conditions to form an inhomogeneity similar to a galaxy
cluster. The characteristic length of the structure will be given by the
parameter $\epsilon=\tau (1+z_i)^{1/2}$, as a sort of Jeans length, where
$z_i$ is the initial redshift and $\tau=\sqrt{T_o/m}$ represents the present
value of the random  mean square  velocity (r.m.s. velocity).

Finally, we face the problem of the validity of the linear approximation. The
validity criterion, working in the longitudinal gauge, cannot be based on
the value of the density contrast, as we have commented above. Instead, it
should be based on the value of the non linear corrections of the evolution
equations. So, in section \ref{VAL} we estimate the first non linear
correction, $\phi^{(1)}$, due to the quadratic terms in the Einstein's
equations, and obtain an upper bound estimation for the quotient $\Gamma=
|\phi^{(1)}/\phi^{(0)}|$. The linear approximation will be considered
suitable if this quotient is small, although the density contrast has reached
a great value. With this validity criterion, an inhomogeneity similar to a
cluster of galaxies could be described with the linear approximation because
we obtain a $\Gamma$ of the order of $10^{-2}$. Notice that, if we take the
well known solution for $p=0$, we obtain $\Gamma$ bigger than unity. So, the
pressure plays a crucial role for the validity of the linear approximation.

\section{The evolution equations}
\label{EV}
We assume in this paper the concepts and notations usual of the $3+1$
formalism of general relativity \citep{YORK}. We shall consider a
perturbation of an Einstein-de Sitter Universe, so we put the metric in the
form:
\begin{equation}
ds^2  =  -\alpha^2 (\phi) dt^2 + \gamma_{ij} dx^i dx^j
\end{equation}
The shift vector $\beta_i $ has been taken null, and the lapse function
$\alpha$ will be choosen conveniently later. We write the tridimensional
metric $\gamma_{ij}$ in terms of a scalar $\phi$ and a trace-less tensor:
\begin{equation}
\label{tridi}
\gamma_{ij}  =  a^2 ((1-2\phi)\delta_{ij}+\sigma_{ij})
\end{equation}
where $a= a(t)$ denotes the scale factor of the Einstein-de Sitter Universe,
and $\sigma_{ij}$ is a tridimensional tensor verifying $ \delta^{mn}
\sigma_{mn}=0$. In the following we shall neglect quadratic terms in the
metric perturbations, $\phi$ and $ \sigma_{ij}$, but those which are
quadratic in its first derivatives (as was explained in the introduction). We
shall need the extrinsic curvature of the surfaces $t=$constant,
$$
K_{ij}:= -\frac{1}{2\alpha}\partial_t \gamma_{ij}= - \frac{a^2}{\alpha}
\{(H(1-2\phi)-\partial_t \phi) \delta_{ij}+ H \sigma_{ij}+
\frac{1}{2}\partial_t \sigma_{ij}\},
$$
being $H = \dot a /a$ the Hubble constant, and the Ricci tensor of the
tridimensional metric $ \gamma_{ij}$:
$$
R_{ij}= (1+2\phi)\phi_{,ij}+3\phi_{,i}\phi_{,j}+((1+2\phi)\Delta \phi+
(\nabla \phi)^2 ) \delta_{ij}-\frac{1}{2}\Delta \sigma_{ij}-\delta^{mn}
\sigma_{(im,mj)}
$$
where the operators $\Delta$ and $\nabla$ are referred to the euclidean
tridimensional metric.

Splitting the energy tensor in parallel and orthogonal components to the
vector field $u= (1/\alpha )\partial_t$,
\begin{equation}
T^{\mu \nu}= \rho u^{\mu}u^{\nu} + p h^{\mu \nu} + q^{\mu}u^{\nu} +
q^{\nu}u^{\mu} + \pi^{\mu\nu},
\end{equation}
one gets the corresponding energy density, f{l}ux of energy and stress tensor.
Then, a Cauchy problem with constraints can be stated in General Relativity
\citep{BRUHAT}. Over a space-like surface $t= t_{i}$ a tridimensional metric
and an extrinsic curvature tensor are suposed to be given. These tensors
evolve in time according to the following equations:
\begin{equation}\label{EVE}
\begin{array}{rcl}
\partial_t \gamma_{ij} &=& -2\alpha K_{ij}  \\
\partial_t K_{ij} &=& -D_iD_j\alpha +\alpha (R_{ij}+tr K K_{ij} -
2K_{ia}K^{a}_{j})+4\pi G\alpha(p-\rho)\gamma_{ij} - 8\pi G\alpha \pi_{ij}
\end{array}
\end{equation}
with $tr K$ representing the tridimensional metric trace of the extrinsic
curvature tensor and $D_i$ the tridimensional covariant derivative. In these
equations the components of the stress tensor, $p$ and $\pi_{ij}$, must be
choosen from the beginning, as we will do below.

The energy density and the f{l}ux of energy are linked by constraint
conditions to the Ricci tensor and the extrinsic curvature:
\begin{equation}\label{COE}
\begin{array}{rcl}
16\pi G\rho &=& (trK)^2-tr(K^2)+R  \\
8\pi G q_i &=& D^a K_{ai}- D_i trK
\end{array}
\end{equation}
where $R$ is the scalar curvature of the tridimensional metric. If one knows
at some initial instant $t_i$ the f{l}ux of energy and the energy density,
one must solve first the constraint equations (\ref{COE}) to determine a
valid set of initial conditions $ \gamma^{*}_{ij}(x) , K^{*}_{ij}(x) $. Then,
the evolution equations (\ref{EVE}) determine the $\gamma_{ij}(t,x),
K_{ij}(t,x)$ for $t > t_i$. Substituting them into the constraint equations
one gets the evolution of the energy density and the flux of energy.

Finally, we shall assume a one-component Universe with a pressure tensor of
the form:
\begin{eqnarray}
\pi_{ij} & = & A [\phi_{,ij}]^t + \pi^{(2)}_{ij}(\phi,t) \label{pi} \\
p & = & p_{B}+  E \Delta \phi + p^{(2)}(\phi,t) \label{pres}
\end{eqnarray}
where $[\phi_{,ij} ]^t$ means trace-less component, $A$ and $E$ are only
functions of time and $p^{(2)}$ and $\pi^{(2)}_{ij}$ are a scalar and a
second order 3-tensor formed with the 3-vectors $\phi_{,i} \ , \
\sigma_{im,m}$ and its first time derivatives respectively. This assumption
is quite general because it allows to consider an ideal gas as well as
solutions of an Einstein-Vlasov problem. Next, we shall develope the
evolution equations taking into account these last expressions.

Let us start splitting the second evolution equation into trace-less and
trace part equations. The trace-less part is:
\begin{equation}
-\frac{a^2}{2}\partial^2_t \sigma_{ij} - \frac{3}{2}a^2 H \partial_t
\sigma_{ij}+\frac{1}{2\alpha} \Delta \sigma_{ij}
+\frac{1}{\alpha} \sigma^t_{(im,mj)}= S_{ij}
\end{equation}
where
\begin{equation}\label{ese}
S_{ij}= (\alpha(1+2\phi)-\alpha' -8\pi G \alpha A) [\phi_{,ij}]^t
 + (-\alpha'' - 2\alpha' +3\alpha) [\phi_{,i}\phi_{,j}]^t- 8\pi G \alpha
 \pi^{(2)}_{ij} .
\end{equation}
An appropriate election of the lapse function $\alpha(\phi)$ can simplify the
problem. Lifshitz used the Gaussian gauge: $\alpha =1$, but this choice has
more than one inconvenience, as we have pointed out at the introduction.
Looking at equation (\ref{ese}) we see that the best choice is to take
$\alpha $ such that the coefficient of the Hessian vanishes. That means to
take $\alpha$ as the solution of the equation:
$$
\alpha(1+2\phi)-\alpha' -8\pi G \alpha A = 0
$$
which is $\alpha = e^{b_1 \phi + \phi^2}$, 
with $b_1 = 1-8 \pi G A$. With this election, the coefficient of $[\phi_{,i}
\phi_{,j}]^t$ in the expression of $S_{ij}$ becomes $-2 + 32 \pi G  A$, and 
the evolution of the trace-less component results:
$$
\partial^2_t \sigma_{ij} + 3  H \partial_t \sigma_{ij}-\frac{1}{a^2} (\Delta
\sigma_{ij}+2\sigma^t_{(im,mj)})= \frac{4}{a^2}(1-16\pi G  A)[ \phi_{,i}
\phi_{,j}]^t+ \frac{16 \pi G(1-8\pi G A)}{a^2} \pi_{ij}^{(2)}
$$

As to the trace component of the evolution equation, it writes down as:
$$
\partial^2_t \phi + 4 H \partial_t \phi - (\frac{8\pi G A} {3 a^2} +
4 \pi G E)\Delta \phi+ \frac{1}{12 a^2} \partial_i \partial_m \sigma_{im}=
\frac{1}{2}(\partial_t \phi)^2 -\frac{1}{6 a^2}(\nabla \phi)^2 + 4 \pi G p^{(2)}
$$   

We have found convenient to introduce the conformal time $\eta$, defined by
$dt=a^2d\eta$. In this time coordinate the expansion factor writes down as
$a(\eta)=a_o\eta^2$ with $a_o$ related to the Hubble constant by $a_o=2/H_o$.
Then,the final form of the evolution equations is:
\begin{equation}\label{GNLE1}
\partial^2_{\eta}\phi + \frac{6}{\eta} \partial_{\eta}\phi -
\frac{4\pi G}{3}(2 A + 3 E a^2)\Delta \phi +
\frac{1}{12} \sigma_{im,im} = \frac{1}{2}(\partial_{\eta} \phi)^2 -
\frac{1}{6}(\nabla \phi)^2+ 4 \pi G a^2  p^{(2)} 
\end{equation} 
\begin{equation}\label{GNLE2}
\partial^2_{\eta} \sigma_{ij} + \frac{4}{\eta}\partial_{\eta}\sigma_{ij} -
\Delta \sigma_{ij}-2[\sigma_{(jm,im)}]^t = 4(1-16\pi G A) [\phi_{,i}
\phi_{,j}]^t + 16 \pi G(1-8\pi G A) \pi_{ij}^{(2)} 
\end{equation}
In addition to these equations, the constraint conditions should be
considered:
\begin{eqnarray}
4\pi G \rho &=& \frac{1}{a^2} \Delta \phi - \frac{3H}{a} \phi_{,\eta}+
\frac{3H^2}{2}(1-2b_1 \phi)+\frac{5}{2a^2}(\nabla\phi)^2 + \frac{3}{2a^2}
(\phi_{,\eta})^2+\frac{1}{4a^2}\sigma_{im,im}  \\
4\pi G q_{i} &=& - b_1 H \phi_{,i}-  \frac{1}{a}\phi_{,\eta i}+ \frac{1}{2a}
\sigma_{im,m\eta}+H \sigma_{im,m}
\end{eqnarray}

So, it remains to give the functions A, E, $p^{(2)}$ and $\pi^{(2)}_{ij}$
appearing in the pressure tensor. As we are interested in the matter
dominated epoch, it is reasonable to consider the one-component Universe as
an ideal gas, with energy tensor $ T_{\mu \nu}= \rho_c w_{\mu} w_{\nu} +
p_c (g_{\mu \nu}+w_{\mu} w_{\nu})$, equation of state $p_c =( T/m) \rho_c$
and temperature evolving as $T= const/a^2$. In the Appendix A we show how
this assumption means to take $A=0$ and $E\approx \tau^2  /4 \pi G  a^4 $,
where $\tau = \sqrt{T_o/m}$ is the r.m.s. veloctiy at the present epoch, and
neglect the second order expressions $p^{(2)}$ and $\pi^{(2)}_{ij}$.

Let us repoduce the complete set of equations:
\begin{eqnarray}
\partial^2_{\eta}\phi + \frac{6}{\eta} \partial_{\eta}\phi -
\frac{\tau^2 }{\eta^4}\Delta \phi +\frac{1}{12} \sigma_{im,im} & = &
\frac{1}{2}(\partial_{\eta} \phi)^2 - \frac{1}{6}(\nabla \phi)^2 \label{NLE1} \\
\partial^2_{\eta} \sigma_{ij} + \frac{4}{\eta}\partial_{\eta}\sigma_{ij} -
\Delta \sigma_{ij} -2[\sigma_{(jm,im)}]^t & = & 4 [\phi_{,i}\phi_{,j}]^t
\label{NLE2}
\end{eqnarray}
\vspace{-0.5cm}
\begin{eqnarray}
\delta & = & \frac{\eta^2}{6} \Delta \phi -  \eta \phi_{,\eta}- 2 \phi +
\frac{\eta^2}{24}\sigma_{im,im} \\
4\pi G q_{i} & = & - H \phi_{,i}- \frac{1}{a}\phi_{,\eta i}+ \frac{1}{2a}
\sigma_{im,m\eta}+H \sigma_{im,m}
\end{eqnarray}
where we have substituted the energy density $\rho$ by the density contrast
$\delta$ using the relation $\delta=(\rho-\rho_{_B})/\rho_{_B}$, with 
$\rho_{_B}=3H^2/8\pi G$ the background energy density.

\subsection{The longitudinal gauge}

The longitudinal gauge, unlike the Gaussian gauge, fixes definitely the
coordinates. This is a well known fact, but let us give here an argument,
which may  be useful for other purposes.

If we start with the Robertson-Walker (R-W) metric in canonic coordinates,
$ds^2 = -d\bar t^2 + a^2(\bar t) \delta_{ij} d\bar x^i d\bar x^j $ and we
introduce new coordinates:
\begin{eqnarray*}
\bar t &=&  t + \varphi( t,  x) \\
\bar x^i &=&  x^i + \xi_i( t,  x)
\end{eqnarray*}
and impose the coordinate conditions $g_{oi} = 0$, one puts the metric in the
form:
\begin{equation}
ds^2 = -(1+2\dot \varphi) d  t^2+ a^2 (t)[ (1+ 2 H \varphi)\delta_{mn}
dx^m dx^n + 2 \xi_{(m,n)} d x^m d x^n]
\end{equation}
with
$$
\xi_m( t,  x)= \frac{1}{2} \psi_{,m} + \zeta_m (x) , \qquad  \psi = 2 \int
\frac{1}{a^2}\varphi dt
$$
being $\dot \varphi$ the time derivative of $\varphi$, and $\zeta_m $
arbitrary functions of the space-like coordinates. Then, the lapse function
and the tridimensional metric are:
\begin{eqnarray*}
\alpha &=&  1+\dot \varphi \\
\gamma^{RW}_{ij} &=&  a^2( t) \left[ (1+2H\varphi  +\frac{1}{3} \Delta \psi+
\frac{2}{3} \zeta_{m,m}) \delta_{ij} + \psi^t_{,ij}+ 2 \zeta^t_{(m,n)}\right]
\end{eqnarray*}
Comparing these expressions with equation (\ref{tridi}) we obtain the
gravitational potential and the trace-less tensor as:
\begin{eqnarray*}
-2 \phi &=& 2H \varphi + \frac{1}{3} \Delta \psi + \frac{2}{3} \zeta_{m,m} \\
\sigma_{mn}&=& \psi^t_{,mn} + 2 \zeta^t_{(m,n)}
\end{eqnarray*}
Consequently, if we choose the longitudinal gauge, i.e. $\alpha = 1+ \phi$,
the function $\varphi$, which define the new time coordinate, should satisfy
the equation:
\begin{equation}
\partial^2_t \varphi + \frac{1}{3a^2}\Delta \varphi + \dot H \varphi + H
\partial_t \varphi =0 .
\end{equation}
Notice that the coefficients of the second time derivatives and of the
laplacian operator in this equation  have the same sign. This makes
impossible to construct a time-like foliation with the function $\varphi$,
apart from the case $\varphi = 0$. So, we can conclude that if we develope a
R-W perturbation in the longitudinal gauge, it is impossible to recover a R-W
space-time in other coordinates. 

However, if one chooses the Gaussian gauge, i.e. $\alpha=1$, the function
$\varphi$ should now satisfy the equation $\dot \varphi=0$, making possible
to introduce new time coordinates. This forces to characterize the gauge
modes, i.e. the false metric perturbations.

\section{The linear evolution equations}\label{LEE}

We shall study now the linear equations:
\begin{eqnarray}
\partial^2_{\eta}\phi + \frac{6}{\eta} \partial_{\eta}\phi -
\frac{\tau^2 }{\eta^4}\Delta \phi + \frac{1}{12} \sigma_{im,im}  & = & 0  \\
 \partial^2_{\eta} \sigma_{ij} + \frac{4}{\eta}\partial_{\eta}\sigma_{ij} -
 \Delta \sigma_{ij} -2[\sigma_{(jm,im)}]^t & = & 0
\end{eqnarray}
\vspace{-0.5cm}
\begin{eqnarray}
\delta & = & \frac{\eta^2}{6} \Delta \phi -  \eta \phi_{,\eta}- 2 \phi
 +\frac{\eta^2}{24}\sigma_{im,im} \label{contrast} \\
4\pi G q_{i} & = &  -  H \phi_{,i}- \frac{1}{a}\phi_{,\eta i}+
\frac{1}{2 a} \sigma_{im,m\eta}+H \sigma_{im,m} \label{flux}
\end{eqnarray}

Firstly, we consider that the trace-less symmetric tensor $\sigma_{ij}$ can
be decomposed \citep{YORK2} in a transverse part $\sigma^T_{ij}$, verifying
$\sigma^T_{im,m}=0$, and a longitudinal part $\sigma^L_{ij}$, with
$\sigma^L_{im,m} = \sigma_{im,m}$. These two components are orthogonal and
evolve independently. Then, one can distinguish three modes in our problem:
the scalar one $\phi$, the transverse tensor $\sigma^T_{ij}$ and the
longitudinal tensor $\sigma^L_{ij}$. The constraint equations show that the
scalar mode is the most important contribution to the density contrast
(\ref{contrast}) and to a rotational-free flux of matter (\ref{flux}); the
longitudinal tensor mode contributes weakly to the density contrast and, what
is more interesting, it is the only possibility of producing a non null
rotational component of the velocity field. Relative to the tensor transverse
mode, it does not contribute nor to the density neither to the energy flux,
in fact it represents the emission of gravitational waves.

Next, given that the scalar and the longitudinal modes are coupled and that
the double divergence of the longitudinal part appears in the evolution of
$\phi$, we can rewrite the equations introducing the scalar $\theta =-(1/12)
\sigma_{im,im}$. In this manner, the evolution of $\sigma_{ij}$ gives the
evolution of $\theta$ and the weak coupling between the scalar mode $\phi$
and the longitudinal one $\theta$ is given by:
\begin{equation}
\label{coupling}
\partial^2_{\eta}\phi + \frac{6}{\eta} \partial_{\eta}\phi -
\frac{\tau^2 }{\eta^4}\Delta \phi  = \theta
\end{equation}
\begin{equation}
\label{theta}
 \partial^2_{\eta} \theta + \frac{4}{\eta}\partial_{\eta}\theta -
 \frac{7}{3} \Delta \theta  = 0
\end{equation}

Both equations are hyperbolic, but there exist a great difference between
them due to the time dependent coefficient $\eta^{-4}$ that appears in the
evolution of $\phi$. This coefficient causes that the characteristic curves
of the first equation do not scape to infinity  as do the charateristics of
$\theta$, because in this case the laplacian operator has a constant
coefficient. This can be seen clearly in the case of spherical symmetry,
where the charateristic  curves $r(\eta)$ of (\ref{coupling}) tends to a
finit limit as the conformal time tends to infinity; while for the variable
$\theta$, the characteristic curves of (\ref{theta}) scape to infinity. This
makes possible that a linear hyperbolic equation might describe the
increasing of density in bounded regions.

Morever, this difference makes  the coupling between the scalar and
longitudinal modes almost irrelevant, because small initial values for
$\theta$ in a small region disperse to infinity.

Finally, we observe  an important difference between our evolution equations
an those in  the Gaussian gauge (lapse function $\alpha = 1$ and shift vector
$\beta = 0$) used by Lifshitz. In our gauge, the evolution equation for the scalar
mode can be reduced to a unique equation for a unique function, while in the
Gaussian gauge the scalar mode is described by two coupled equations.

In the next section we shall find the general solution in the coordinate
space of the evolution equation for the scalar mode, neglecting the coupling
with $\theta$ or assuming  $\theta(\eta_{i},x)= \partial_{\eta} \theta
(\eta_{i},x)=0$.

\section{The  Cauchy problem of the linear evolution equations}\label{CPLE}

From equation (\ref{coupling}) and in the case of null initial conditions for
$\theta $, we can  consider the following initial value problem:
\begin{equation}\label{CP}
 \begin{array}{rcl}
  \partial^2_{\eta}\phi + \frac{\textstyle 6}{\textstyle\eta} \partial_{\eta}
  \phi - \frac{\displaystyle\tau^2}{\displaystyle\eta^4}\Delta \phi  & = & 0
  \\ \phi(\eta_{i},x)  =  \phi_{i}(x) \ , \ \partial_{\eta}\phi(\eta_{i},x) &
   = &  \phi'_{i}(x)
\end{array}
\end{equation}
with $\phi_{i}(x)$ and  $\phi'_{i}(x)$ two arbitrary functions. We shall
solve this problem using the method of Fourier transforms.  

Firstly, let us denote by $\hat\phi(\eta,s)$, with $s \in \mathbb{R}^3$, the
Fourier transform of $\phi(\eta,x)$ with respect to the spatial coordinates.
In the Fourier space, equation (\ref{CP}) transforms into an initial value
problem for an ordinary differential equation:
\begin{equation}\label{CPO}
 \begin{array}{rcl}
  \partial^2_{\eta}\hat\phi + \frac{\displaystyle 6}{\displaystyle\eta}
  \partial_{\eta}\hat\phi + \frac{\displaystyle\tau^2}{\displaystyle\eta^4}
  s\cdot s \hat\phi  & = & 0 \\
  \hat\phi(\eta_{i},s)  =  \hat\phi_{i}(s) \,\
  \partial_{\eta}\hat\phi(\eta_{i},s) & = &  \hat\phi'_{i}(s)
 \end{array}
\end{equation}

So, the first task is that of constructing  a system of fundamental solutions,
which consists on two solutions $\hat\phi_1(\eta, s)$, $\hat\phi_2(\eta, s)$
that satisfy the initial conditions:
\begin{eqnarray*}
 \hat\phi_1(\eta_{i},s)  = 1 \qquad , \qquad
 \partial_{\eta}\hat\phi_1(\eta_{i},s)=0  \\
 \hat\phi_2(\eta_{i},s)  = 0 \qquad , \qquad
 \partial_{\eta}\hat\phi_2(\eta_{i},s) = 1
\end{eqnarray*}
These fundamental solutions can be obtained using complex Laplace transforms
\citep{SMIRNOV}, having the following result:
\begin{equation}\label{funda1}
\hat\phi_1(\eta,s) = \frac{3}{\epsilon^3}\left( \frac{\sin k g}{k^3}- g
 \frac{\cos k g}{k^2} \right ) + \frac{\eta_{i}(3\eta - \eta_i)}{\epsilon \eta^2}\frac{\sin k g}{k}+
 \frac{\eta^2_{i}}{\eta^2}\cos kg
\end{equation}
\begin{equation}\label{funda2}
\begin{array}{rcl}
 \hat\phi_2(\eta, s) & = & \frac{\textstyle 9 \eta_{i}}{\textstyle\epsilon^5} \left (
 \frac{\textstyle\sin k g}{\textstyle k^5}- g \frac{\textstyle\cos k g}{\textstyle k^4}-g^2 \frac{\textstyle\sin kg}{\textstyle 3 k^3}
 \right )+ \\ \\
 & + & \frac{\textstyle 3 \eta_{i}^2}{\textstyle\epsilon^3 \eta}\left ( \frac{\textstyle\sin k g}{\textstyle k^3}- g
 \frac{\textstyle\cos k g}{\textstyle k^2}\right ) + \frac{\textstyle\eta_{i}^3}{\textstyle\epsilon \eta^2}
 \frac{\textstyle\sin k g}{\textstyle k}
\end{array}
\end{equation}
where $k$ stands for the modulus of $s$, $k=\sqrt{s\cdot s}$, $\epsilon =
\tau/\eta_{i}$, and $g= \epsilon (1-\frac{\eta_i}{\eta})$. Then, the solution
of (\ref{CPO}), in the Fourier space, is expressed in terms of the fundamental
system as:
\begin{equation}
\hat\phi(\eta, s)=\hat\phi_{i}(s)\hat\phi_1(\eta, s)+ \hat\phi'_{i}(s)
\hat\phi_2(\eta, s) .
\end{equation}

The next task is to obtain the Green's functions $Q_1(\eta,x)$ and
$Q_2(\eta,x)$, defined as the inverse Fourier transform of the fundamental
system $\{ \hat \phi_{1}(\eta,s), \;  \hat \phi_{2}(\eta,s) \}$. As we show
in Appendix B, these Green functions are:
\begin{eqnarray}
Q_1(\eta, x) & = & \frac{3}{4\pi\epsilon^3}H(g-r)+(\frac{3\eta_{i}}{\epsilon
\eta}- \frac{\eta^2_{i}}{\epsilon \eta^2})\frac{\delta_D(r-g)}{4 \pi g}+
\frac{\eta^2_{i}}{4\pi \eta^2}\partial_{g}\left (\frac{\delta_D(r-g)}{g}
\right ) \label{Green1} \\
Q_2(\eta, x) & = & \left ( \frac{3\eta_i}{8\pi \epsilon^5} (g^2- r^2) +
\frac{3\eta^2_i}{4\pi \eta \epsilon^3}\right ) H(g-r)
+ \frac{\eta^3_{i}}{4\pi\epsilon \eta^2}\frac{\delta_D(r-g)}{g} \label{Green2}
\end{eqnarray}
Therefore, the solution of (\ref{CP}) in the coordinate space is expressed as
the convolution product of the Green's functions with the initial conditions:
\begin{equation}\label{cauchysolution}
\phi(\eta,x)= Q_1(\eta,x)*\phi_{i}(x)+ Q_2(\eta,x)* \phi'_{i}(x)
\end{equation}
where $*$ stands for the convolution product with respect to the spatial
coordinates. Looking at the Green's functions we can observe that the
solution tends rapidly to a limit when the conformal time tends to infinity,
this limit has a simple expression:
\begin{equation}
\label{ACS}
\phi(\infty,x)= \frac{3}{4\pi \epsilon^3 }\int_{\mid x-\xi \mid < \epsilon }
\phi_i(\xi) d \xi + \frac{3 \eta_i}{8 \pi \epsilon^5}
\int_{\mid x-\xi \mid < \epsilon }(\epsilon^2 -\mid x-\xi \mid^2)
\phi'_i(\xi) d \xi
\end{equation}
as tridimensional integrals of the initial conditions. The $\epsilon$
parameter in expression ({\ref{ACS}) can be also written as $\epsilon =
\tau \sqrt{ 1+z_i} $, with $\tau $ the r.m.s. velocity of the matter
component at the present epoch and $z_i$ the initial redshift (recall the
relation $1+z=1/\eta^2 $ in an Einstein-de Sitter Universe). This parameter
will be crucial to the study of evolution because it will fix the
characteristic length of the evolved structures as a sort of Jeans length.

In the case of spherical symmetry the convolutions reduce to unidimensional
integrals, whose expressions are obtained in the Appendix B. They will be
used in the sequel to discuss some examples.

\section{The evolution of fluctuations with non null thermal motions}
\label{EFNN}

Having got the general solution of the linear initial value problem, the
constraint equations (\ref{contrast}) and (\ref{flux}) determine the density
contrast and the flux of matter  as simple functionals of the metric
perturbations. In the linear approximation we have that these equations for
$\delta$ and $q_m$ reduce to:
\begin{eqnarray} \label{delta}
\delta(\eta,x) &=&  \frac{\eta^2}{6} \Delta \phi -2 \phi  -  \eta \phi'  \\
4 \pi G a  q_m &=& - a H \phi_{,m} - \phi_{,\eta m}
\end{eqnarray}
where we only have to substitute the expression (\ref{cauchysolution}) of the
solution $\phi$. Notice that in the case of an ideal gas, the flux of energy
in the longitudinal gauge represents the macroscopic velocity: $q_m= \rho V_m$.
In particular we are interested in the velocity norm, whose expression is:
\begin{equation} \label{velocity}
\mid V(\eta,x) \mid_{\gamma} = \frac{1}{1+\delta} \mid \nabla (\frac{\eta}{3}
\phi + \frac{\eta^2}{6} \phi_{, \eta}) \mid .
\end{equation}

Let us study these expressions when the general solution (\ref{cauchysolution})
is substituted. We shall assume that at some initial redshift $z_i$ we know
the initial conditions for the potential $\phi_i(x)$ and its first time
derivative $ \phi'_i(x) $. We shall begin with a qualitative description of
the evolution of the density contrast based on the following reduced
expression for $\delta$:
\begin{eqnarray}
\label{cda}
\delta(\eta,x) & \approx & \frac{\eta^2}{6}( Q_1(\infty,x) * \Delta \phi_i(x)
 + Q_2(\infty,x)* \phi'_i(x)) \\
 Q_1(\infty,x) &=& \frac{3}{4 \pi \epsilon^3} H(\epsilon - r) \\
 Q_2(\infty,x) &=&  \frac{3 \eta_i}{8 \pi \epsilon^5}(\epsilon^2 - r^2)
 H(\epsilon - r)
\end{eqnarray}
where we have only considered the laplacian term  neglecting the $\phi$ and
$\phi'$ contributions and we have also taken the asymptotic values for the
Green's functions. The idea is to obtain  $L^p$ estimations of the
convolutions using the H\H{o}lder inequalities \citep{Hormander}. In this
case we have enough with the relations $ \parallel f * g \parallel_{\infty}
\leq \parallel f \parallel_1 \parallel g \parallel_{\infty}$, where
$\parallel f \parallel_{\infty}$ means $ sup \mid f \mid $, and
$\parallel f \parallel_{1}$ means $ \int \mid f \mid dx $. We obtain in this
way, for $\eta >> \eta_i $, two upper bounds for $\delta$:
\begin{equation}
\label{c1}
\parallel \delta(\eta,x)\parallel_{\infty} \leq  \frac{\eta^2}{6}\Big (
\parallel Q_1 (\infty,x) \parallel_{\infty} \  \  \parallel \Delta \phi_i(x)
\parallel_1 + \parallel Q_2 (\infty,x) \parallel_{\infty} \  \  \parallel
\Delta \phi'_i(x) \parallel_1 \Big)
\end{equation}
\begin{equation}
\label{c2}
\parallel \delta(\eta,x)\parallel_{\infty} \leq  \frac{\eta^2}{6}\Big(
\parallel Q_1 (\infty,x) \parallel_{1} \  \  \parallel \Delta \phi_i(x)
\parallel_{\infty} + \parallel Q_2 (\infty,x) \parallel_{1} \  \  \parallel
\Delta \phi'_i(x) \parallel_{\infty} \Big)
\end{equation}
This will allow to reach the main conclusions with little calculations,
treating separately each one of our two degrees of freedom $\phi_i(x)$ and
$\phi'_i(x)$. In the subsections we shall give more details in numerical
examples where we shall assume spherical symmetry. To do it we need to fix
the parameters of the problem. We have three parameters related by the
condition $\epsilon =\tau \sqrt{1+z_i}$. As we have mentioned above the final
characteristic length of a structure will depend on the value of $\epsilon$,
so if we want to discuss galaxy clusters it will be  convenient to take
$\epsilon \approx 1Mpc/a_o $, $a_o = 6000 h^{-1}Mpc $, which is of the order
of an Abel's radius for this value of $\epsilon$. And assuming $z_i = 5000$
we obtain $\tau \approx 10^{-6} $ for  the  r.m.s. velocity at the present
epoch. Notice that this value can be supported by hot particles as neutrinos
with non null mass.

In these examples, to identify the final structure  as something similar to a
galaxy cluster, we shall estimate the total mass  at the present epoch $\eta
= 1 $ and inside an Abel's radius $r_a = 1.5 h^{-1} Mpc$ as a function of the
amplitude $A$ and the initial characteristic length $R$:
\begin{equation}\label{mass}
 M(<r_a)= 4 \pi  2.7 \times 10^{11} M_{\odot} \int_{0}^{r_a}
 \delta(r,1,R,A)dr
\end{equation}
Recall that a typical galaxy cluster has $ M(<r_a)\approx 3 h^{-1}\times
10^{14} M_{\odot}$.

In the  subsections we examine the possibility of generating structures
similar to a galaxy cluster starting from reasonable initial conditions for
the potential, that is, such that the initial density contrast and the
macroscopic velocity be small. We shall also require that in both examples
the gravitational potential at the decoupling of matter and radiation be
smaller than $6\times 10^{-5}$.

\subsection{ Initial conditions of the form $\phi_i \neq 0$, $\phi'_i = 0 $}
\label{A}

Let us start studying the case of $\phi_i \neq 0$ and $\phi'_i = 0$. In this
case, from equations (\ref{delta}) and (\ref{velocity}), one gets the initial
density contrast and the initial macroscopic velocity as: $\delta_i(x)
\approx \frac{\eta_i^2}{6} \Delta \phi_i(x)$ and $\mid V(\eta_i,x)
\mid_{\gamma} \approx \frac{\eta_i}{3} \mid\nabla \phi_i \mid $.  Let us
consider an initial potential of the form  $\phi_i(x) = -A (1+r^2/R^2)^{-1/2}$,
in which we have two parameters, the amplitude $A$ and the characteristic
length $R$. The laplacian of this function is $\Delta\phi_i = 6
B(1+r^2 /R^2)^{-5/2}$, where $B$ is given by the relation $A= 2BR^2$ and is
proportional to the initial central density contrast, $\delta_i(0)= \eta_i^2
B$. Under these considerations, from equations (\ref{c1}) and (\ref{c2}) we
have that the final density contrast is upper bounded by:
$$
\parallel \delta(\eta,x) \parallel_{\infty} \leq minor \ \ of \ \ \left[
 \frac{\eta^2}{\eta_i^2} \frac{R^3}{\epsilon^3} \delta_i(0) \ , \
 \frac{\eta^2}{\eta_i^2} \delta_i(0)\right]
$$

So, one distisguishes two cases:
\begin{itemize}
 \item If $R < \epsilon $ we have $\parallel \delta(\eta,x) \parallel \leq
       (\eta/\eta_i)^2 (R/\epsilon)^3 \delta_i(0) $. In this case the density
       contrast decreases until the instant $\eta_{*} $, for which $(\eta_{*}/
       \eta_i)^2 (R/\epsilon)^3 = 1$, and then begin to increase with a power
       law. This possibility was stated first by Gilbert \citep{GILBERT} using
       a Newtonian approximation to the Einstein equations.
 \item If $R>\epsilon$ the object grows from the very beginning, tending to a
       power law: $\parallel \delta(\eta,x) \parallel \leq  (\eta/\eta_i)^2
       \delta_i(0)$.
\end{itemize}

Getting on with the example we will assign numerical values to the free
parameters $A$ and $R$. We shall choose them such that the total mass inside
an Abel sphere of radius $r_a = 1.5 h^{-1} Mpc$ be about $3 h^{-1}\times
10^{14}$ solar masses and, at the same time, keeping bounded the
gravitational potential by  $\mid \phi(\eta,x)\mid < 6\times 10^{-5}$. For
example, taking $A= 2.8 \times 10^{-5}$ we can obtain  the mass $M(<r_a)$ as
a function of the characteristic length $R$, whose graph is given in Figure
\ref{f1}.

\begin{center}
\begin{figure}[!ht] 
  \begin{center}
\includegraphics{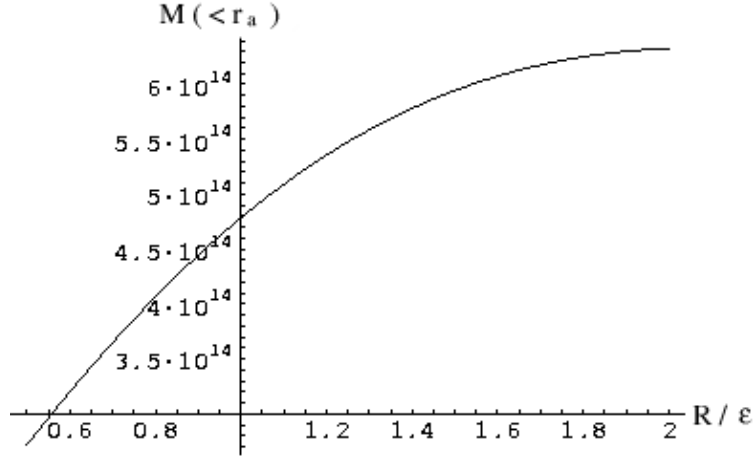}
  \end{center}
  \caption{\label{f1} The mass $M (<r_a)$ inside an Abel's radius ($ r_a =
           1.5 h^{-1} Mpc$ ) at the preset epoch as a function of the initial
           characteristic length $R$, given by equation (\ref{mass}) when we
           evolve the density contrast taking into account only the first
           degree of freedom, i.e., having as initial conditions $\phi_i = -A
           (1+r^2/R^2)^{-1/2}$ and $\phi'_i = 0 $, with $A= 2.8\times 10^{-5}$.
           We assume the values $\epsilon =  1 Mpc/a_o  \ , \ a_o = 6000 h^{-1}
           Mpc \ ,\ h=0.5 $ and $z_i= 5000$ for the $\epsilon $-parameter and
           the initial redshift. The r.m.s. velocity at the present epoch
           parameter takes the value $\tau = 1.2 \times  10^{-6}$.}
 \end{figure} 
\end{center}
This figure shows that we can choose $R= 1.6 \epsilon$ to obtain the mass of
a typical galaxy cluster. With this election, equations (\ref{velocity}) and
(\ref{cda}) give the evolution of the velocity and the density contrast,
whose graphs at the initial and final time are shown  in  Figure \ref{f2}.

\begin{center}
\begin{figure}[!ht]
  \begin{center}
 \includegraphics[width=16.2 cm]{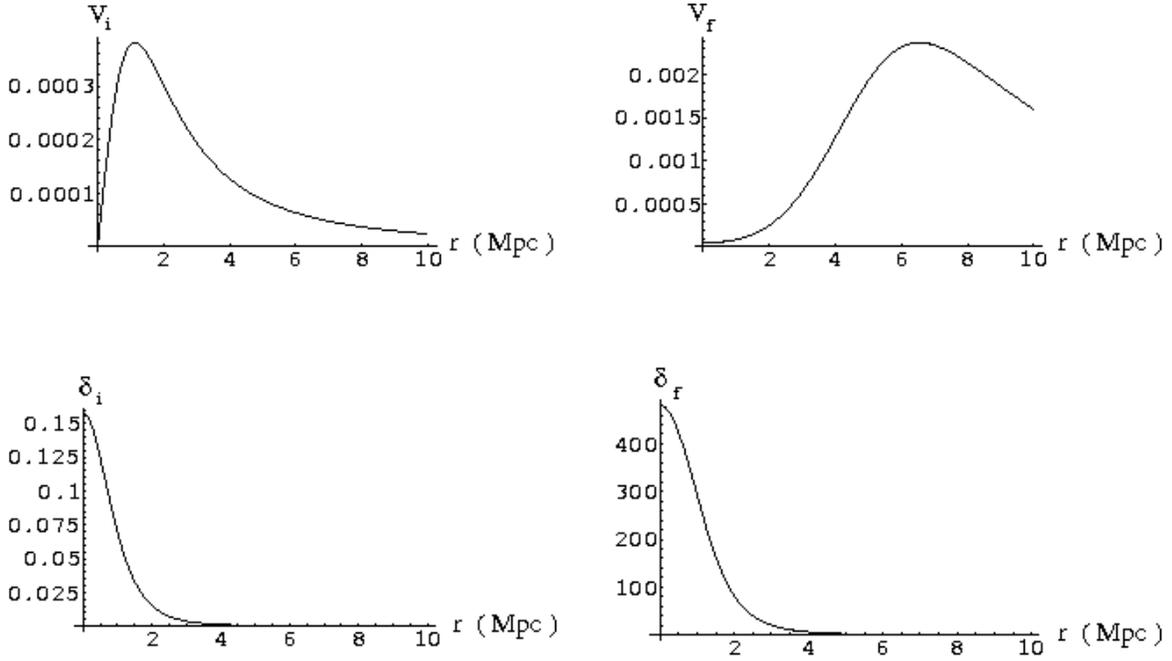} 
  \end{center}
  \caption{\label{f2} Initial (left figures) and present (right figures)
           values for the density contrast and the macroscopic velocity,
           evolving the first degree of freedom under the same conditions as
           in Figure 1 and with $R=1.6\epsilon$. The mass inside an Abel's
           radius $r_a = 1.5 h^{-1}Mpc $ at the present epoch is about
           $6 \times 10^{14} M_{\odot}$.}
 \end{figure}
\end{center}

In Figure \ref{f3} we show the evolution of the central value of the density
contrast in the case where $R<\epsilon$ to illustrate the bouncing of
inhomogeneities with typical length $R$ smaller  than $\epsilon$, as we have
described above.

\begin{center}
\begin{figure}[!ht]
  \begin{center}
\includegraphics{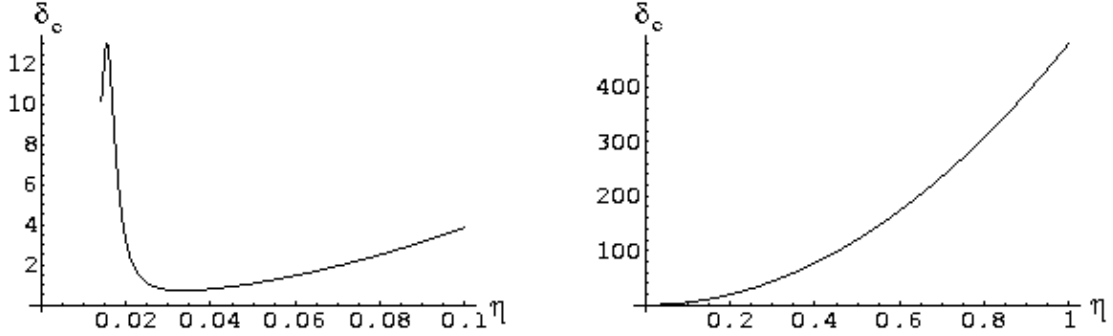}
  \end{center}
  \caption{\label{f3}The central density contrast obtained evolving the first
           degree of freedom. The right figure shows the evolution under the
           same conditions as in Figure 1 and with $R=1.6\epsilon$. The left
           figure shows the bouncing produced on the central density contrast
           when we take an initial characteristic length $R= 0.2 \epsilon$.
           Then we can see that the $\epsilon$ parameter plays the role of a
           Jeans length.}
 \end{figure}
\end{center}

As conclusion of this subsection we can say that the evolution of Einstein's
equations with initial conditions of the form $\{ \phi_i(x)\neq 0,\phi'_i(x)
=0 \}$ is equivalent to the evolution goberned by Newton's equations
\citep{GILBERT} with initial conditions $ \{\delta_i(x), V_i(x)\} $ given at
the beginning of the subsection. The second example will lead us to a quite
different conclusion.

\subsection{Initial conditions of the form $\phi_i= 0$, $\phi'_i \neq 0 $}
\label{B}

Next we are going to study the case where  $\phi_i= 0$ and  $\phi'_i \neq 0 $.
In this case, the  initial  density contrast and the macroscopic velocity are not given by  the initial potential but by 
 its  initial first  time derivative in the form:  $ \delta_i(x) =  - \eta_i \phi'_i $  and  
$\mid  V(\eta_i,x) \mid_{\gamma} =  \frac{\eta_i^2}{6} \mid  \nabla \phi'_i \mid $. 
Now, we shall consider an initial condition  of the form   $\phi'_i(x) = -A (1+r^2/R^2)^{-1/2} $, with $A$ and $R$ two free parameters. Defining $A= 2BR^2$ we can write  $\Delta\phi'_i  = 6  B(1+r^2 /R^2)^{-5/2}$ having that the final density contrast is  upper bounded by:
$$
\parallel \delta(\eta,x) \parallel_{\infty} \leq   minor \  \ of \ \   \left[
 \frac{1}{2}\eta^2 \eta_i  \frac{R^3}{\epsilon^3} B \ , \   \frac{1}{5}\eta^2 \eta_i B \right] .
$$

Let us remark the main difference with the previous case. If we consider here $R \geq \epsilon $, the final density contrast will be bounded by $(1/5) \eta_i B $, but now $B$ is not constrained to be small  because  the initial density contrast and the macroscopic velocity as  functions of $B$ and $R$ are given by:
\begin{eqnarray}
\label{New2}
\delta_i(r) &=& \frac{2\eta_i BR^2}{ (1+r^2/R^2)^{1/2}}  \\
\mid V(\eta_i,r) \mid_{\gamma}  &=& \frac{\eta^2_i Br}{3(1+r^2/R^2)^{3/2}} \label{New3}
\end{eqnarray}
having that for objects much smaller than the horizon at the present epoch ($R=1Mpc/a_o $ with  $a_o=6000 h^{-1} Mpc $ implies  $R \sim  10^{-4}$),    $\eta_i BR^2$  can  be  small even when $B >> 1$. Therefore, the second degree of freedom allows to reach great values of the density contrast starting with very small initial density contrast.

As in  the previous subsection, we have to take  values for $A$ and $ R$
such that the total mass inside an Abel sphere of radius
$r_a = 1.5 h^{-1} Mpc $ be of the order of $3 h^{-1}\times  10^{14}$ solar
masses  and, at the same time, keeping bounded  the gravitational potential
by  $\mid \phi(\eta,x)\mid < 6\times 10^{-5}$.  For example, taking
$A= 3 \times 10^{-2}$ we can determine $M(<r_a) $ as a function of $R$, whose
graph is represented in Figure \ref{f4}.

\begin{center}
\begin{figure}[!ht]
  \begin{center}
\includegraphics{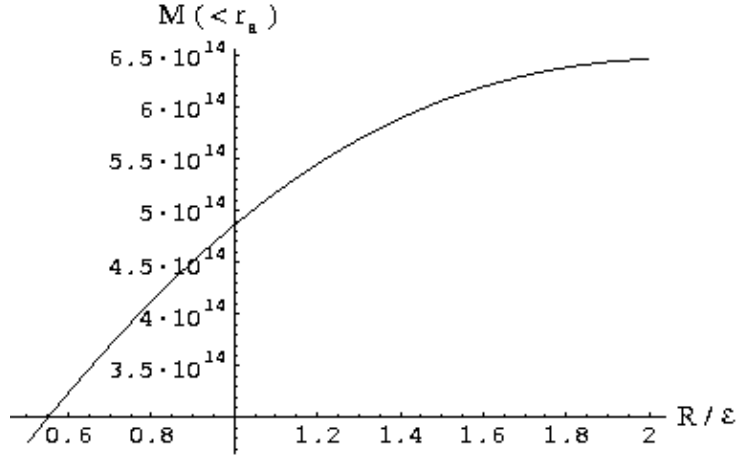}
  \end{center}
  \caption{\label{f4}The mass $M (<r_a)$ inside an Abel's radius ($ r_a = 
           1.5 h^{-1} Mpc$ ) at the preset epoch as a function of the initial 
           characteristic length $R$, given by equation (\ref{mass}) when we 
           evolve density contrast taking into account only the second degree 
           of freedom, i.e., having as initial conditions $\phi_i = 0$ and 
           $\phi'_i = -A(1+r^2/R^2)^{-1/2}$, with $A=10^{-2}$. We assume the 
           same values for $\epsilon$, $z_i$, $\tau$ and $h$ as in Figure 1.}
 \end{figure}
\end{center}
From this picture we obtain that a good value for the characteristic length
is $R= 1.6 \epsilon $. These values of the parameters allow us to evolve the
density contrast and the velocity, whose evolution is represented in Figure
\ref{f5}.

\begin{center}
\begin{figure}[!ht]
  \begin{center}
\includegraphics{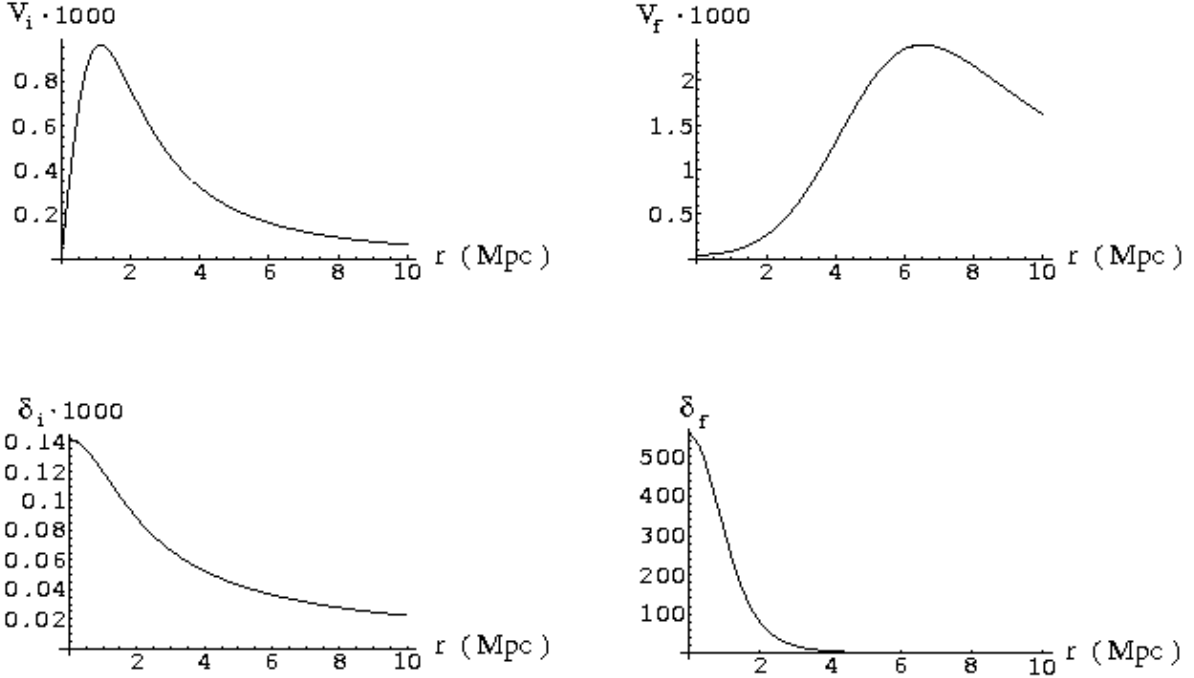}
  \end{center}
  \caption{\label{f5} Initial (left figures) and present (right figures) values 
           for the density contrast and the macroscopic velocity, evolving the 
           second degree of freedom with the choice of parameters done in Figure 
           4 and with $R=1.6\epsilon$. The mass inside an Abel's radius $r_a = 
           1.5 h^{-1}Mpc $ at the present epoch is $6 \times 10^{14} M_{\odot}$.}
 \end{figure}
\end{center}

Then, we can conclude that the second degree of freedom, taking an appropriate 
value for $\phi'_i$, allows the formation of great structures starting from very 
small initial values for the density contrast.  This case has no Newtonian analogue 
because now we have significant initial time derivatives of the gravitational 
potential. In other words, a Newtonian evolution starting with initial density 
contrast and macroscopic velocities as given by expressions (\ref{New2}) and 
(\ref{New3}) will produce a structure with a very small final density contrast at 
the present epoch.

Finally, let us remark that the second degree of freedom can be described 
geometrically as follows: the initial surface $\eta = \eta_i$  has null intrinsic 
curvature (null laplacian  of $\phi_i$) and highly inhomogeneous extrinsic 
curvature (great space derivatives of $\phi'_i$).

\section{On  the validity of the linear approximation}
\label{VAL}

In this section we come back to the non linear equations (\ref{NLE1}) and (\ref{NLE2}) 
in order to study the validity of the linear approximation. Introducing the function  
$\theta$ as in section \ref{LEE}, we obtain a coupled system of evolution equations 
for the couple of functions $(\phi, \theta )$:
\begin{eqnarray*}
\partial^2_{\eta}\phi + \frac{6}{\eta} \partial_{\eta}\phi - \frac{\tau^2 }{\eta^4}
\Delta \phi &=& \theta + \frac{1}{2}(\partial_{\eta} \phi)^2 - \frac{1}{6}(\nabla \phi)^2 \\
\partial^2_{\eta} \theta + \frac{4}{\eta}\partial_{\eta}\theta -\frac{7}{3} 
\Delta \theta  &=& -\frac{1}{3}\partial_a \partial_b [\phi_{,a}\phi_{,b}]^t
\end{eqnarray*}
with initial conditions $\phi(\eta_i,x)=  \phi_i(x) ,\, \partial_{\eta}\phi(\eta_i,x)
= \phi'_i(x) ,\, \theta(\eta_i,0)=0,$ and $\partial_{\eta}\theta(\eta_i,x)=0 $. This 
is a semilinear hyperbolic initial value problem. In the Courant-Hilbert's book  
\citep{COHI} the unicity of solutions of this kind of problem is shown by means of the 
convergence of iterations. This supports the fact of considering the first iteration as 
criterion for validity of the linear approximation. In the following we shall focus 
on the reduced equation:
\begin{equation}
\partial^2_{\eta}\phi + \frac{6}{\eta} \partial_{\eta}\phi - \frac{\tau^2 }{\eta^4}
\Delta \phi  =  - \frac{1}{6}(\nabla \phi)^2
\end{equation}
because  it contains the essentials of the problem. Let us denote by $\phi^{(0)}$ the 
solution (\ref{cauchysolution}) to the linearized equation and by $\phi^{(1)}$ the 
first non linear correction, namely the solution of 
\begin{equation}
\partial^2_{\eta}\phi^{(1)} + \frac{6}{\eta} \partial_{\eta}\phi^{(1)} - 
\frac{\tau^2 }{\eta^4}\Delta \phi^{(1)}  = - \frac{1}{6}(\nabla \phi^{(0)})^2
\end{equation}
with  null initial conditions. Using the Fourier transform method, the problem reduces 
to an ordinary differential equation:
\begin{equation}
\label{first}
\partial^2_{\eta}\hat\phi^{(1)} + \frac{6}{\eta} \partial_{\eta}\hat\phi^{(1)} 
+ \frac{\tau^2 }{\eta^4} s\cdot s \hat\phi^{(1)}   =  L^{(0)}(\eta , k)
\end{equation}
where $L^{(0)}(\eta,k)$ stands for the Fourier transform of the quadratic term $- \frac{1}{6}(\nabla \phi^{(0)})^2$, which 
is easily solved by the method of  constants variation. The fundamental solutions $\{\hat\phi_1(\eta, s),\hat\phi_2(\eta, s)\}$ of the homogeneous equation  were 
obtained in section \ref{CPLE}, see expressions (\ref{funda1}) and (\ref{funda2}). Then,  the solution of (\ref{first}) can be expressed by means of integrals:
$$
\hat \phi^{(1)}(\eta, k ) = \frac{\eta_i}{\epsilon}\left ( \hat \phi_2 \int_0^{g(\eta)} (1-\frac{g}{\epsilon})^4 
\hat \phi_1  L^{(0)}  dg - \hat \phi_1 \int_0^{g(\eta)} (1-\frac{g}{\epsilon})^4 
\hat \phi_2  L^{(0)} dg \right )
$$
and in the coordinate space it results:
\begin{equation}
\begin{array}{rcl}
\phi^{(1)}(\eta, x ) & = & \frac{\textstyle\eta_i}{\textstyle\epsilon}\left ( Q_2* {\displaystyle\int}_0^{g(\eta)} (1-\frac{\textstyle g}{\textstyle\epsilon})^4 
Q_1 *  L^{(0)}(g,x)  dg - \right. \\ & - & \left.
 Q_1 * {\displaystyle\int}_0^{g(\eta)} (1-\frac{\textstyle g}{\textstyle\epsilon})^4 
Q_2 *  L^{(0)}(g,x) dg \right )
\end{array}
\end{equation}
being $Q_1$ and $Q_2$ the Green's functions given by (\ref{Green1}) and (\ref{Green2}).
To decide about the validity of the linear approximation we need to compare $ \mid \phi^{(1)}(\eta ,x) \mid $ with $ \mid \phi^{(0)}(\eta ,x) \mid $ . This is not easy to do directly, but  using $L^p$ norms we shall get an upper bound for  the first non linear correction  which will be enough to discuss the problem. Then, we have:
$$
\mid \phi^{(1)}(\eta, x )\mid \ \ \leq \ \ \frac{\eta_i}{\epsilon}\left ( \mid Q_2* \int_0^{g(\eta)} (1-\frac{g}{\epsilon})^4 
Q_1 *  L^{(0)}(g,x)  dg \mid + \ldots  \right )
$$
where dots means the same expression but interchanging $Q_2$ for $Q_1$. As before we are going to use  the H\H older inequalities, in particular $ \parallel f*g \parallel_{\infty} \  \ \leq  \ \ 
\parallel f \parallel_1 \ \  \parallel g \parallel_{\infty}$  and  $ \parallel f*g \parallel_{2} \  \ \leq  \ \ 
\parallel f \parallel_1 \ \  \parallel g \parallel_{2}$, where $\parallel g \parallel_{2}$ means $\int \mid g \mid^2 dx $. We get in this  way:
$$
\parallel  \phi^{(1)}(\infty, x) \parallel_{\infty} \ \ \leq \ \  \frac{\eta_i}{\epsilon}\left ( \parallel  Q_2(\infty ,x) \parallel_1  \int_0^{\epsilon} (1-\frac{g}{\epsilon})^4 
\parallel Q_1(g,x) \parallel_{\infty}   \parallel L^{(0)}(g,x) \parallel_1   dg  + \ldots  \right )
$$

In the sequel we shall obtain this upper bound for  the numerical  example studied in  subsection \ref{B}  corresponding to the initial conditions of the form $\phi_i(x)= 0 $ and $ \phi'_i \neq 0 $. 
 So,  we  write  $\phi^{(0)}(\eta, x) = Q_2(\eta,x)* \phi'_i(x)$  and   get the estimation $\parallel L^{(0)}(g,x) \parallel_{_1} \ \ \leq  \ \ \frac{1}{6} \parallel Q_2(g,x) \parallel^2_{_1} \sum^3_{a=1} \parallel \nabla_a \phi'_i(x) \parallel^2_{_2} $. Substituting this  into the previous equation we obtain: 
\begin{equation}\label{primera}
\parallel \phi^{(1)}(\infty , x )\parallel_{\infty} \ \ \leq \ \ \frac{1}{6} G( \eta_i , \epsilon ) \sum^3_{a=1} \parallel \nabla_a \phi'_i(x) \parallel^2_2
\end{equation}
with
\begin{eqnarray*}    
G(\eta_i ,\epsilon ) & = & \frac{\eta_i}{\epsilon}\left ( \parallel  Q_2(\infty,x) \parallel_1  \int_0^{\epsilon} \left(1-\frac{g}
{\epsilon}\right)^4 \parallel Q_1 \parallel_{\infty} \ \  \parallel  Q_2 \parallel^2_1 dg + \right. \\
 & & \left. \quad + \parallel Q_1(\infty,x) \parallel_1  \int_0^{\epsilon} \left(1-\frac{g}{\epsilon}\right)^4 
\parallel Q_2 \parallel_{\infty} \ \   \parallel  Q_2 \parallel^2_1 dg \right )
\end{eqnarray*}  
Evaluating the corresponding  $L^p$ norms we  have:
\begin{small}
$$
G(\eta_i, \epsilon) = \frac{3 \eta_i^4}{4 \pi \epsilon^3} \left (  \frac{1}{5}  \int_0^{1} (1-y)^4(\frac{y^5}{5} + y^3 - y^4)^2 dy + 
 \int_0^{1} (1-y)^4(\frac{y^5}{5} + y^3 - y^4)^2(\frac{y^2}{2} +1 -y) dy \right )
$$
\end{small}
which reduces, once calculated the integrals, to:
\begin{equation}
G(\eta_i, \epsilon)= 10^{-5} \frac{27 \eta_i^4}{4 \pi \epsilon^3}
\end{equation}

Then, in the case considered in the previous subsection \ref{B}, a simple calculation gives  $\sum^3_{a=1} \parallel \nabla_a \phi'_i(x) \parallel^2_2 \  \ = \  \ 3 \pi^2 B^2 R^5 $. Substituting these results  into equation (\ref{primera}),  we get that the upper bound $ {\cal C}^{(1)}$ to the first non linear correction is:
\begin{equation}
\label{quota}
 \mid \phi^{(1)}(\infty, x ) \mid  \ \ \leq  \ \  \parallel \phi^{(1)}(\infty , x )\parallel_{\infty} \ \ < \ \  {\cal C}^{(1)} =  10^{-4}\eta_i^4 \frac{R^3}{\epsilon^3} B^2 R^2 
\end{equation}

Now, we have to compare this bound with the norm of the linear solution $\phi^{(0)}$. To do it  we form the quotient $\Gamma = {\cal C }^{(1)}/\mid \phi^{(0)}(\infty,0) \mid $.
Given that for this example we have spherical symmetry, we can use the unidimensional integrals of Appendix B to calculate the modulus of  $\phi^{(0)}$. Then, $\Gamma$  expresses  as a function of the initial characteristic length $R$, which is represented in Figure  \ref{f6}.  

\begin{center}
\begin{figure}[!ht]
  \begin{center}
\includegraphics{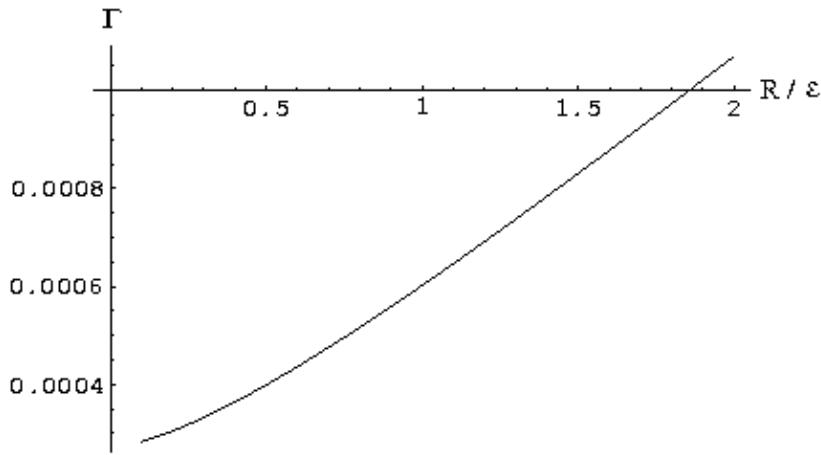}
  \end{center}
  \caption{\label{f6}Validity of the linear aproximation of the evolution of the second degree of freedom showed in Figure
           5. The function $\Gamma$ is an upperbound, at the present epoch,  of the ratio of the first non linear 
           correction and the linear solution for the gravitational potential.}
 \end{figure}
\end{center}

As we can see in this figure, the quotient $\Gamma $ is below $10^{-2}$ in the range $ R \leq 2 \epsilon $,  then the relation  $\mid \phi^{(1)}(\infty,x) \mid  < 0.01 \mid \phi^{(0)}(\infty,x) \mid $ is verified. So, as we have choosen $R=1.6 \epsilon$, we can neglect the first non linear correction and consequently, the linear approximation is an accurate description of this  problem even having reached a great final density contrast.  

Therefore, we can conclude that a  thermal  velocity $\tau$ of the order of $10^{-6}$ at the present epoch makes possible to follow with the linear approximation  the formation of an inhomogeneity similar to a galaxy cluster. We have seen also that the linear approximation  comes into problems with smaller values for the thermal velocity. 
  
\section{Conclusions}
 
In the current theory on evolution of perturbations, the matter dominated
epoch is considered as a fluid with null pressure, and the evolution is
described using the linear approximation until the density contrast becomes
of the order of unity. However, as we have shown in this paper, if pressure
is properly considered, the evolution with the linear approximation can be
extended to values of the density contrast bigger than unity. We have assumed
an isotropic pressure of the form $p= (a_o^2 T_o/ma^2(t))\rho$, which
corresponds to an ideal gas with uniform temperature, and we have used
reasonable values for the temperature. Concretely, in our examples we have
taken a random mean square velocity of the order of $\sqrt{T_o/m}
\approx 10^{-6}$.

In the following we summarize the main steps we have followed to get these
conclusions:
\begin{enumerate}
 \item We have stated a Cauchy problem using sistematically the 3+1
       formalisme of General Relativity and neglecting quadratic terms in the
       metric perturbation. But, given that for inhomogeneities at scales of
       a few Mpc the spatial derivatives of the potential are much bigger
       than the potential, we have kept the quadratic terms in its first
       derivatives. These non linear corrections will only be used to study
       the validity of the linear approximation (see point 3). As usual, the
       coordinates are fixed by choosing the lapse function $\alpha$ and the
       shift vector $\beta$. We have put $\beta=0$, and $\alpha=
       e^{b_1 \phi + \phi^2}$ has been taken in order to simplify the
       evolution equations. When linearized, this choice of coordinates is
       called the longitudinal gauge.
 \item We have obtained the  solution of the linearized Cauchy problem for a
       one-component Universe in the matter dominated epoch and assuming an
       ideal gas equation of state $p= (a_o^2 T_o/ma^2(t))\rho$. We have
       expressed this solution in terms of convolution integrals of the
       initial conditions, which in our case are the initial potential
       $\phi_i(x)$ and its first time derivative $\phi'_i(x)$. We have also
       studied how to obtain a density inhomogeneity similar to a cluster of
       galaxies, i.e., how to get a total mass inside an Abel radius of the
       order of $3 h^{-1}\times 10^{14}$ solar masses. We have considered
       separately both degrees of freedom obtaining in both cases an
       inhomogeneity similar to a galaxy cluster at the present epoch. But
       the initial density contrast and the initial macroscopic velocity in
       each case are very different:
  \begin{enumerate}
   \item With initial conditions of the type $\{  \phi_i(x) \neq 0 \ , \
         \phi'_i(x)=0 \} $, see subsection \ref{A}, one gets a cluster of
         galaxies starting at redshift $z_i = 5000$. These inital conditions
         correspond to an initial density contrast of $\delta_i \sim 0.1$
         (which becomes about $0.5$ at the recombination redshift), and a
         macroscopic velocity of $\mid V_i \mid \sim 3 \times 10^{-3}$. The
         results of this case can be also obtained with a Newtonian analysis
         starting with the same density contrast and macroscopic velocity.
   \item With initial conditions of the type $\{ \phi_i(x)=0 \ , \ \phi'_i(x)
         \neq 0 \}$, see subsectio \ref{B}, one gets a galaxy cluster starting
         at the same redshift $z_i = 5000 $. But now one has a very small
         initial density contrast $\delta_i \sim 0.0001$, and a similar
         macroscopic velocity, $\mid V_i \mid \sim 8 \times 10^{-3}$. Unlike
         the previous case, this evolution has no Newtonian analogue.
  \end{enumerate}
  
       The second degree of freedom, which is currently forgotten, may
       rapidly produce inhomogeneities similar to galaxy clusters starting
       from faint initial density contrast. On the contrary, the first
       degree of freedom needs an excessive initial density contrast.
 \item We have estimated the first non linear correction to the linear
       approximation and used it as the criterion of validity. Our first
       results seem quite interesting: assuming at the present epoch a
       thermal velocity of the order of $\tau \approx 10^{-6}$, the quotient
       between the first correction to the gravitational potential
       $\phi^{(1)}$ and the linear solution $\phi^{(0)}$ is small than
       $10^{-2}$. Therefore, we can conclude that the linear approximation is
       an accurate description of the formation of a big structure if the
       effect of the pressure is not neglected.

\end{enumerate}

\appendix

\section{Description of an ideal gas in the longitudinal gauge}

An ideal gas in the matter dominated epoch with four-velocity $w$ has a perfect fluid energy tensor  $ T_{\mu \nu}= \rho_c w_{\mu} w_{\nu} + p_c (g_{\mu \nu}+w_{\mu} w_{\nu})$, with  equation of state $p_c=T\rho_c/m$ being   $T= T_o a_o^2/a^2$, where $T_o$ is the temperature at the present epoch. The four-velocity $w$ is related to the four-velocity $u=(1/\alpha)\partial_t$ by $w =\gamma (V)( u + V)) $ where $V$ is the macroscopic velocity of the  Einstein-de Sitter perturbation in the longitudinal gauge, which is given by:
$$
V_i = \frac{1}{4 \pi G\rho}(-  H \phi_{,i}- \frac{1}{a}\phi_{,\eta i}+ \frac{1}{2a} \sigma_{im,m\eta}+H \sigma_{im,m})
$$
where $\rho$ is the energy density for the observer $u$. Using this relation, the energy tensor transforms into  $T_{\mu \nu}= \rho u_{\mu} u_{\nu} + p (g_{\mu \nu}+u_{\mu} u_{\nu})+ q_{\mu} u_{\nu}+q_{\nu} u_{\mu}+ \Pi_{\mu \nu}$, where now $\rho$, $p$ and  $\Pi_{\mu \nu}$ are quantities referred to the observer $u$ and are given by:
\begin{eqnarray}
\rho = \rho_c + O(V^2) \\
p  = p_c + \frac{1}{a^2}\rho V^2 +O(p_c V^2) \\
\Pi_{ij}= \rho [V_i V_j]^t + O(\rho V^4) \label{pig}
\end{eqnarray}

Given the equation of state and the relation $\rho_c = \rho_{_B}(1+ \delta )$ we can write
 $p_c= T\rho_{_B}/m + \delta \rho_{_B} T/m$, which allows  us to get:
$$
p = \frac{T}{m}\rho_{_B} + \frac{\tau^2}{4 \pi G a^2 } \left(\frac{1}{a^2}\Delta \phi - \frac{3H}{a}\phi_{\eta}-3H^2 \phi \right)+ \qquad
$$
$$
+ \frac{1}{18\pi G a^2 H^2(1+\delta)}\left(H^2 (\nabla \phi)^2 + \frac{1}{a^2}(\nabla \phi')^2+ 2\frac{H}{a}\nabla \phi \cdot \nabla \phi' \right)
$$
where we have also  introduced the r.m.s. velocity at the present epoch $ \tau^2= (T_o /m)$. From this expression and comparing with equation (\ref{pres}), we identify: 
\begin{eqnarray}
E&=&\frac{  \tau^2 }{ 4 \pi G a^4 }  \\
p^{(2)}&=& \frac{1}{18\pi G a^2 H^2(1+\delta)}\left(H^2 (\nabla \phi)^2 + \frac{1}{a^2}(\nabla \phi')^2+ 2\frac{H}{a}\nabla \phi \cdot \nabla \phi' \right)
\end{eqnarray}

As for the anisotropic pressures, in the same way, equation (\ref{pig}) gives:
\begin{equation}
\Pi_{ij}= \frac{1}{6 \pi G H^2(1+\delta)} \left ( H^2 [\phi_i \phi_j]^t + \frac{1}{a^2} [\phi'_i \phi'_j]^t  +\frac{H}{a} [\phi'_i \phi_j + \phi_i \phi'_j]^t  \right)
\end{equation}
and comparing with (\ref{pi}) we  obtain $A=0$, and $\pi^{(2)}_{ij} = \Pi_{ij}$.

Let us write the non linear evolution equations (\ref{GNLE1}) and (\ref{GNLE2}) as follows:
\begin{eqnarray*}
{\cal L}^1(\phi, \sigma) & = & \frac{1}{2}(\partial_{\eta} \phi)^2 - \frac{1}{6}(\nabla \phi)^2 + 4 \pi G a^2  p^{(2)} \\
{\cal L}^2(\sigma) & = & 4 (1-16\pi G A) [\phi_{,i} \phi_{,j}]^t + 16 \pi G(1-8\pi G A) \pi^{(2)}_{ij}
\end{eqnarray*}
where we have used a compact notation for the first members of the equations. These expressions, substituting $A$, $p^{(2)}$ and $\pi^{(2)}_{ij}$, transform in:
\begin{eqnarray}
{\cal L}^1(\phi, \sigma) =\frac{1}{2}(\partial_{\eta} \phi)^2 - \frac{1}{6} (\nabla \phi)^2 + \frac{2}{3 H^2(1+\delta)}\left(H^2 (\nabla \phi)^2 + \frac{1}{a^2}(\nabla \phi')^2+ 2\frac{H}{a}\nabla \phi \cdot \nabla \phi' \right) \\
{\cal L}^2(\sigma) = 4  [\phi_{,i} \phi_{,j}]^t +  \frac{8}{3  H^2(1+\delta)} \left ( H^2 [\phi_i \phi_j]^t + \frac{1}{a^2} [\phi'_i \phi'_j]^t  +\frac{H}{a} [\phi'_i \phi_j + \phi_i \phi'_j]^t  \right)
\end{eqnarray}
And taking into account that the density contrast $\delta$ will have a great value in the structures  we are interested on, the terms where it appears can be neglected. 

Let us to point out that keeping these terms only would  produce small corrections to the upper bounds estimations for the first nonlinear correction to the linear approximation as obtained in section \ref{VAL}.

\section{Obtaining the Green functions}

In this appendix we are going to summarize the process to obtain the Green's
functions associated to the general solution of our initial value Cauchy
problem (\ref{CP}). This general solution in the Fourier space has the form:
$$
\hat{\phi}(\eta,s)=\hat{\phi}_i(s)\hat{\phi}_1(\eta,s) +
\hat{\phi}_i^{\prime}(s)\hat{\phi}_2(\eta,s).
$$
being $\hat{\phi}_i(s)$ and $\hat{\phi}_i^{\prime}(s)$ the Fourier transform
of the initial conditions and $\hat{\phi}_1(\eta,s)$ and
$\hat{\phi}_2(\eta,s)$ the fundamental solutions given by (\ref{funda1}) and (\ref{funda2}).
In order to have the general solution in the coordinate space we need to make
the corresponding inverse Fourier transforms with respect to the spatial
coordinates. Taking into account that, in this general development, the  initial conditions are arbitrary functions of the
spatial coordinates,
the best way to give this general solution will be using the convolution
product between functions (denoted by $*$) with respect to the spatial
coordinates. That is, the properties of the inverse Fourier transforms and of
the convolution product allow us to write the solution as:
$$
\phi(\eta,x)=Q_1(\eta,x)*\phi_i(x)+ Q_2(\eta,x)*\phi_i^{\prime}(x)
$$
where $Q_1(\eta,x)$ and $Q_2(\eta,x)$ are the Green functions, that is the
inverse Fourier transforms of the fundamental solutions $\hat{\phi}_1(\eta,s)$ and
$\hat{\phi}_2^{\prime}(\eta,s)$ respectively.
Then, we need to calculate a few inverse Fourier transforms to obtain the
Green functions.

Firstly, we consider that the fundamental functions (\ref{funda1}) and (\ref{funda2}) can also be 
written in the following form:
\begin{equation}\label{fiuno}
\hat{\phi}_1(\eta,s)=\frac{3}{\epsilon^3}{\cal D}_g\left(\frac{\sin kg}{k^3}
\right) +\frac{\eta_i(3\eta-\eta_{i})}{\epsilon \eta^2}\frac{\sin kg}{k} +
\frac{\eta_{i}^2}{\eta^2}\partial_g\left(\frac{\sin kg}{k}\right)
\end{equation}
\begin{equation}\label{fidos}
\hat{\phi}_2(\eta,s)  =\frac{9\eta_i}{\epsilon^5}\left( {\cal D}_g 
+\frac{1}{3}g^2\partial_g^2 \right) \frac{\sin kg}{k^5} 
+ \frac{3\eta_i^2}{\epsilon^3\eta}
{\cal D}_g\left(\frac{\sin kg}{k^3}\right)+ \frac{\eta_i^3}{\epsilon \eta^2}
\frac{\sin kg}{k}
\end{equation}

where $g=\epsilon\left( 1-\frac{\textstyle\eta_i}{\textstyle\eta}\right)$,
$\epsilon=\tau/\eta_i$, $k=\sqrt{s\cdot s}$ and being the operator
$D_g(f)=f-g\partial_g f$. This form simplifies the number of inverse Fourier
tranforms that we have to obtain. In fact, we shall only need a pair of
well--known inverse Fourier transforms as we shall see below.

Let us remind the definition of the inverse Fourier transform of a
function with respect to the spatial coordinates, that is:
$$
{\cal F}^{-1}[W(s,\eta)]=\frac{1}{(2\pi)^3}\int_{{\mathbb R}^3}
           e^{-is\cdot x}W(s,\eta)ds .
$$
With this definition, we have that the inverse Fourier transform of the
function $(\sin kg)/k$ is known and has the following general form:
\begin{equation}\label{inverse}
{\cal F}^{-1}\left[\frac{\sin k\lambda}{k}\right]=\frac{1}{4\pi\lambda}
\left\{ \delta_D(\lambda-r)H(\lambda)+\delta_D(\lambda+r)H(-\lambda)\right\}
\end{equation}
where $\delta_D(x)$ represents the Dirac delta distribution and
$H(\lambda)$ is the Heaviside unity function.
On the other hand, we also have the general expression \citep{guelf}:
\begin{equation}\label{gamma}
{\cal F}^{-1}[k^{-\lambda-n}]=\frac{\Gamma(\frac{-\lambda}{2})r^{\lambda}}
{2^{\lambda+n}\Gamma(\frac{\lambda+n}{2})\pi^{3/2}}
\end{equation}
where $n$ denotes the dimension of the space where are realized the inverse
Fourier transforms and $\Gamma(\cdot)$ represents the Gamma function. These
two expressions will allow us to calculate all the inverse Fourier transforms
involved in the Green's functions.

Having a look to the fundamental solutions (\ref{fiuno}) and (\ref{fidos}) we
can see that we need the inverse Fourier tranforms of functions of the form
$(\sin kg)/k^p$, which can be obtained from (\ref{inverse}) and (\ref{gamma})
using the convolution product in the following way:
$$
{\cal F}^{-1}\left[\frac{\sin k\lambda}{k^p}\right]=
{\cal F}^{-1}\left[\frac{\sin k\lambda}{k}\right]*
{\cal F}^{-1}\left[k^{-(p-1)}\right] \, .
$$

As the convolution product in general is given by the expression:
\begin{equation}\label{conv}
(f*g)(x)=\int_{{\mathbb R}^n}f(t)g(x-t)dt
\end{equation}
it results that the two inverse Fourier tranforms needed are expressed in
general as:
$$
{\cal F}^{-1}\left[\frac{\sin k\lambda}{k^3}\right]=\frac{1}{4\pi}
\left\{ \left[ H(\lambda-r)+\frac{\lambda}{r}H(r-\lambda)\right]H(\lambda)
-\left[ H(-\lambda-r)-\frac{\lambda}{r}H(\lambda+r)\right]H(-\lambda)\right\}
$$
$$
{\cal F}^{-1}\left[\frac{\sin k\lambda}{k^5}\right]=\frac{-1}{24\pi r}
\Big\{ \left[ r(r^2+3\lambda^2)H(\lambda-r)+\lambda(\lambda^2+3r^2)
H(r-\lambda)\right]H(\lambda)-  \qquad\qquad
$$
$$
 -\left[ r(r^2+3\lambda^2)H(-\lambda-r)-
\lambda(\lambda^2+3r^2)H(\lambda+r)\right]H(-\lambda)\Big\}
$$

The corresponding operators acting over these expressions will give us that
the Green functions have the form:
$$
Q_1(\eta,x)=\frac{3}{4\pi\epsilon^3}H(g-r)+\frac{\eta_i(3\eta-\eta_{i})}
{4\pi\epsilon \eta^2}\frac{\delta_D(r-g)}{g} +
\frac{\eta_{i}^2}{4\pi\eta^2}\partial_g\left(\frac{\delta_D(r-g)}{g}\right)
$$
$$
Q_2(\eta,x)=\left(\frac{3\eta_i}{8\pi\epsilon^5}(g^2-r^2)+ \frac{3\eta_i^2}
{4\pi\epsilon^3\eta}\right)H(g-r)+ \frac{\eta_i^3}{4\pi\epsilon \eta^2}
\frac{\delta_D(r-g)}{g}
$$
To obtain these expressions we have considered that $g$ is always a positive
number (that is, $H(g)=1$ and $H(-g)=0$).

Finally, as we have said above, the general solution of the initial value
Cauchy problem in the real space is given by the convolution product between
the Green functions and the initial conditions, that is:
$$
\phi(\eta,x)=Q_1(\eta,x)*\phi_i(x)+ Q_2(\eta,x)*\phi_i^{\prime}(x) .
$$
This expression reduces to unidimensional integrals in the case of spherical
symmetry. That is, if we consider the initial conditions $\phi_i(x)=f_1(r)$
and $\phi_i^{\prime}(x)=f_2(r)$ as functions depending only on the radial
coordinate $r$, then the corresponding convolution product, defined by
(\ref{conv}), is written as unidimensional integrals as we show in the following.

To make clear the expressions we can consider firstly the case when
$f_1(r)\neq0$ and $f_2(r)=0$. In this case the final gravitational potential
will be written as:
$$
\phi(\eta,r)= \frac{3}{2\epsilon^3}\left\{ 2\int_0^{g-r}q^2f_1(q)\, dq\;
H(g-r)+ \frac{1}{2r} \int_{|r-g|}^{r+g} q f_1(q) \Big(2rq+K(r,q,g)\Big)dq
\right\}+
$$

$$
+\frac{\eta_i g(3\eta-\eta_i)}{2\eta^2\epsilon}
\int_{-1}^1 f_1(\sqrt{r^2+g^2-2rgx})dx +
\frac{\eta_i^2}{2\eta^2}\int_{-1}^1 f_1(\sqrt{r^2+g^2-2rgx})dx +
$$

$$
+ \frac{g\eta_i^2}{2\eta^2}\int_{-1}^1
\frac{g-rx}{\sqrt{r^2+g^2-2rgx}}f_1^\prime
(\sqrt{r^2+g^2-2rgx)} dx 
$$

\noindent
where $K(r,q,g)=g^2-r^2-q^2$.

On the other hand, the case in which we have $f_1(r)=0$ and $f_2(r)\neq 0$
we will have that the final evolution of $\phi(\eta, x)$ is:
$$
\phi(\eta,r)= \frac{3\eta_i}{4\epsilon^5}\left\{ 2\int_0^{g-r}q^2f_2(q)
K(r,q,g) dq  H(g-r) - \frac{1}{4r} \int_{|r-g|}^{r+g}q
 f_2(q)\Big(K(r,q,g)^2+ 4r^2q^2\Big) dq +  \right.
$$
$$ \left. 
+\frac{1}{2r}\int_{|r-g|}^{r+g} q K(r,q,g) f_2(q)\left( 2rq+ K(r,q,g)
\right) dq \right\} +
\frac{\eta_i^3g}{2\eta^2\epsilon}\int_{-1}^{1}
 f_2(\sqrt{r^2+g^2-2rgx}) dx +
$$
$$
+ \frac{3\eta_i^2}{2\eta\epsilon^2}\left\{ 2\int_0^{g-r} q^2f_2(q)dq H(g-r)+
 \frac{1}{2r}\int_{|r-g|}^{r+g} qf_2(q)\left( 2rq+K(r,q,g)\right) dq
 \right\}
$$

\begin{acknowledgments}
 The authors would like to thank D. S\'aez, V. Quilis, J.A. Morales, J.J. Ferrando  for their helpful 
discussions. This work has been partially supported by the Spanish MCyT, project number AYA 2000--2045. 
\end{acknowledgments}

% now the references. delete or change fake bibitem. delete next three
%   lines and directly read in your .bbl file if you use bibtex.

\end{document}